\documentclass[%
 amsmath,amssymb]{revtex4-2}
  \usepackage{xcolor}
 \usepackage{hyperref}
\hypersetup{
    colorlinks=true,
    linkcolor=orange,
    filecolor=magenta,      
    urlcolor=cyan,
    citecolor=purple,
    pdftitle={Overleaf Example},
    pdfpagemode=FullScreen,
    }

\usepackage{graphicx}
\usepackage{dcolumn}
\usepackage{bm}
\usepackage{siunitx} 
\usepackage{gensymb} 
\usepackage{mhchem} 
\newcommand{\comment}[1]{}

\begin{document}


\title{Electron-to-photon noise transfer in mid-infrared semiconductor lasers}

\author{Irene La Penna}
\email[]{lapenna@lens.unifi.it}

\affiliation{LENS - European Laboratory for Non-Linear Spectroscopy, Via Carrara 1 - 50019 Sesto Fiorentino FI, Italy}
\affiliation{CNR-INO - Istituto Nazionale di Ottica, Largo Fermi, 6 - 50125 Firenze FI, Italy}
\author{Tecla Gabbrielli}
\affiliation{LENS - European Laboratory for Non-Linear Spectroscopy, Via Carrara 1 - 50019 Sesto Fiorentino FI, Italy}
\affiliation{CNR-INO - Istituto Nazionale di Ottica, Largo Fermi, 6 - 50125 Firenze FI, Italy}
\author{Borislav Hinkov}
\affiliation{ SAL- Silicon Austria Labs, Europastraße 12 - 9524 Villach, Austria}
\author{Robert Weih}
\affiliation{Nanoplus Nanosystems and Technologies GmbH - Gleimershäuser Strasse 10 - 98617 Meiningen, Germany}
\author{Naota Akikusa}
\affiliation{SSD advanced innovation headquarters, Hamamatsu Photonics K.K., Shizuoka 431-2103, Japan}
\author{Lorenzo Mischi}
\author{Alessio Montori}
\affiliation{ppqSense, Viale Ariosto 492B - 50019 Sesto Fiorentino FI}
\author{Simone Borri}
\author{Luigi Consolino} 
\author{Francesco Cappelli} 
\author{Paolo De Natale}
\affiliation{LENS - European Laboratory for Non-Linear Spectroscopy, Via Carrara 1 - 50019 Sesto Fiorentino FI, Italy}
\affiliation{CNR-INO - Istituto Nazionale di Ottica, Largo Fermi, 6 - 50125 Firenze FI, Italy}

\date{\today}

\begin{abstract}
Noise characteristics of state-of-the art light sources are crucial parameters in understanding their limitations towards quantum applications. 
This work describes a method to study the electrical noise transfer of current driver sources to the intensity noise of mid-infrared emission by commercial quantum and interband cascade lasers (QCLs and ICLs, respectively). A current driver with sub-shot electrical noise in a specific frequency range (up to 10 dB below the shot noise level) was developed for this purpose. This enables testing the performance of mid-infrared lasers when driven via such a quiet pump source. By using this novel current driver, we identify the fundamental noise of a QCL and an ICL, that is the laser intensity noise resulting solely from the internal dynamics of the laser under test. The proposed methodology allows us to retrieve the noise transfer function from current to light, showing that the main limitations in observing the quantum properties of the emitted photons come from laser excess noise and poor matching between laser and detection system in terms of bandwidth and optical power. From the analysis of the measured parameters, we highlight current technological limitations and suggest which key features should be optimized in mid-infrared systems for matching the performance required by quantum applications. (228 words)
\end{abstract}

\maketitle


\section{Introduction}
The control and reduction of frequency and intensity noise of coherent-light sources are of broad interest for different applications, such as metrology, interferometry, and spectroscopy, which require highly stable light sources \cite{borri_high-precision_2019, miller_design_1987, juretzka_intensity_2015}. In the unceasing quest to improve the sensitivity and precision of mid-infrared (MIR) spectroscopy techniques, the frequency noise of the most commonly used compact light sources, which are quantum cascade lasers (QCLs) and interband cascade lasers (ICLs), has been extensively studied. Previous works include theoretical modeling and validation~\cite{borri_frequency-noise_2011,bartalini_measuring_2011, borri_unveiling_2020, wang_rate_2018,  deng_rate_2020} as well as development of novel techniques aimed at narrowing the laser linewidth~\cite{taubman_frequency_2002,barbieri_phase-locking_2009,cappelli_subkilohertz_2012,argence_quantum_2015,de_cumis_microcavity-stabilized_2016,shehzad_10_2019,deng_narrow_2020,consolino_controlling_2021}. However, less effort has been put in the reduction of intensity noise, with only a few literature entries accounting for its study in MIR lasers~\cite{gentsy_intensity_2005,simos_intensity_2014,juretzka_intensity_2015,deng_rate_2020,kim_intensity_2024,marschick_mid-infrared_2024,gabbrielli_shot-noise-limited_2025}. This is despite being a key quantity for several topics, including sensing and communication~\cite{corrias_analog_2022,seminara_characterization_2022}. Such applications would significantly benefit from using low-intensity-noise optical sources, which would increase the signal-to-noise ratio drastically. Recently, the development and commercialization of novel technology in the MIR is opening the possibility of accessing squeezed states of light in this spectral region~\cite{gabbrielli_mid-infrared_2021}. As a matter of fact, the emission of potentially quantum-correlated twin beams was predicted to occur in QCLs emitting harmonic frequency combs~\cite{gabbrielli_intensity_2022,popp_modeling_2024} since these devices exhibit high intrinsic nonlinearities characterizing their active medium~\cite{mansuripur_single-mode_2016}. In particular, their high third-order nonlinearity triggers intermodal Four-Wave-Mixing (FWM) processes among the generated emission modes~\cite{friedli_four-wave_2013,gabbrielli_intensity_2022}.
The first successful attempt to set up a balanced detector capable of detecting non-classical states of MIR light (i.e. signals below the shot-noise level of reference) is reported in ref.~\cite{gabbrielli_mid-infrared_2021}. This detector has been applied to test state-of-the-art MIR laser sources~\cite{marschick_mid-infrared_2024,chomet_anti-correlation_2023,gabbrielli_shot-noise-limited_2025} and correlated twin modes emitted by QCLs~\cite{gabbrielli_intensity_2022}. This configuration allowed the observation of correlations at the classical level. However, the excess intensity noise in harmonic combs emitted by QCLs prevented the revelation of correlations at or below the shot-noise level. 
Recently, ICLs have also emerged as interesting candidates for possible non-classical light generation~\cite{marschick_mid-infrared_2024, gabbrielli_shot-noise-limited_2025}. The possibility of generating frequency combs resulting from nonlinear effects in QCLs is similarly present in ICLs~\cite{schwarz_monolithic_2019,sterczewski_near-infrared_2019}. This makes ICLs, in principle, also capable of producing squeezed states of light. 
\\
Triggered by these possible applications, this work describes and assesses a method to study the reduction of the intensity noise of MIR lasers. This is done by following the path suggested by Yamamoto et al.~\cite{yamamoto_amplitude_1986}, i.e., pumping the laser with a quiet pump to reduce the intensity fluctuations of the emitted light. The hypothesis of a transfer of the statistics from the pump process to the emitted light was first theoretically described by Golubev and Sokolov in 1984~\cite{golubev_photon_1984}. In electrically pumped devices, the electron statistics of the bias current is transferred to the emitted light proportionally to the quantum efficiency of the optical emission process, possibly yielding a sub-Poissonian emission. To date, the generation of sub-shot-noise radiation has been observed employing high-efficiency, near-infrared semiconductor lasers pumped with a low-noise current~\cite{yamamoto_photon_1992,marin_squeezing_1995,zhang_quantum_1995,giacobino_quantum_1996,kaiser_amplitude-squeezed_2001,ding_observation_2024}. 
As novelty, this well-established technique is applied here to MIR laser sources via a custom-made ultra-low-noise (sub-shot-noise level) current driver. For this purpose, we measure the electron-to-photon noise transfer in two different MIR cascade devices to compare their specific performance. 
\\
To sum up, this work aims at assessing the possibility of intensity noise reduction in MIR cascade lasers by using a sub-shot-noise drive current. This is done by studying and quantifying the contribution of electrical current fluctuations to the intensity noise of radiation emitted by QCLs and ICLs. In addition, we evaluate the suitability of the tested lasers and the detector to perform quantum measurements in the MIR in view of future improvements.

\section{Methods and discussion}
\subsection{Device characterization}

Typically, noise analysis requires preliminary measurements to qualify both the lasers to be characterized and the employed detector, as well as an optimized noise-suppressed current driver, as detailed below.

\paragraph{Lasers}
Concerning the devices under test, a characterization of light power-current-voltage (LIV) curves and emission spectra has been performed. The candidates for this comparison are two MIR lasers whose active region is based on multiple quantum wells but relies on different light-generation mechanisms. These lasers, whose device structure is explained in detail in appendix \ref{app:lasers}, are in particular: i) a ridge-waveguide DFB ICL whose emission wavelength is approximately \SI{4.5}{\micro m}, and ii) a commercial ridge-waveguide DFB QCL emitting at \SI{4.57}{\micro m}. In the following, we will refer to them simply as ICL and QCL, respectively.
\\ The optical spectra of the lasers are presented and discussed in detail in appendix \ref{app:optical_spectra} (see Fig. \ref{fig:OS}), to ensure that both lasers have single-mode emission at all investigated operating conditions. This ensures that no extra noise and irregularities in the power vs. current curves are introduced by the presence of multiple modes. 
\begin{figure}[htb!]
\centering\includegraphics[width=0.8\linewidth, keepaspectratio]{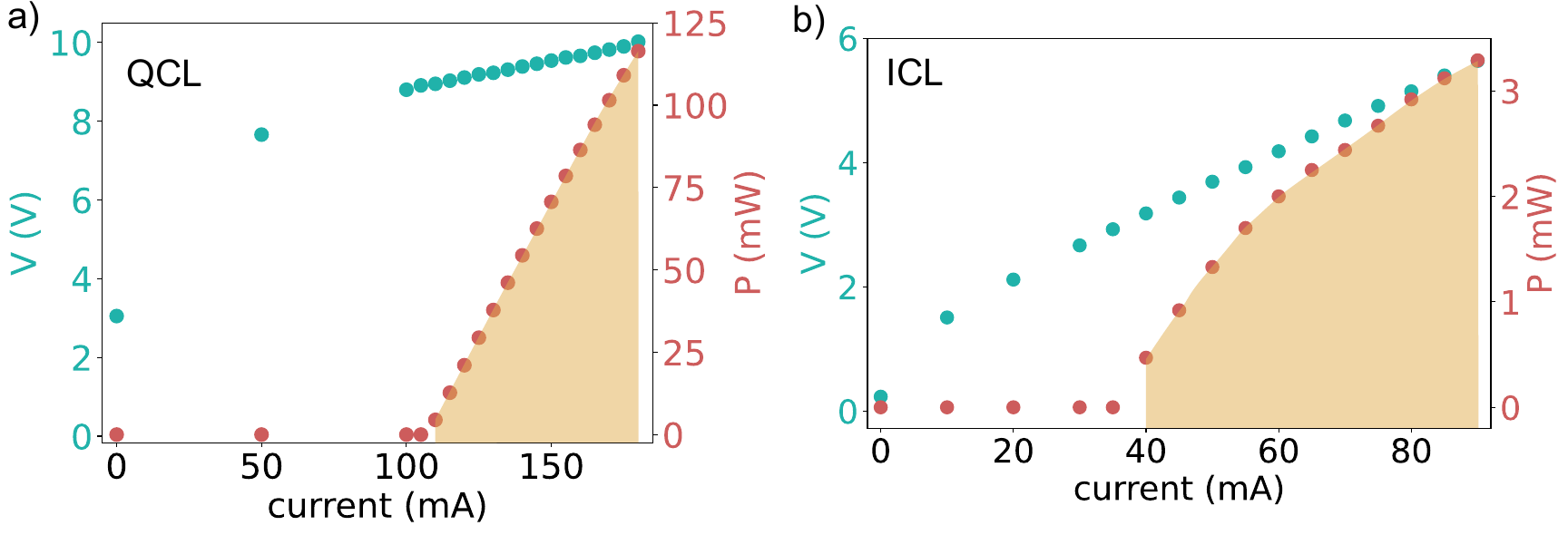}
\caption{CW LIV curves of the two lasers. The turquoise circles represent voltage, while the pink circles represent optical power, both as functions of the applied current. (a) refers to the QCL at T = $\mathrm{20\degree C}$, (b) to the ICL at T = $\mathrm{20\degree C}$. In both plots, the orange-shaded region represents the range of bias current that has been investigated in the subsequent noise analysis. The orange area highlights the profile of the LI curve, showing that the slope of the line is not constant in the region under test for the ICL. The laser temperature is the same as the one used later in the noise analysis.}
\label{fig:LIV}
\end{figure}
The LIV curves of the two lasers operating in continuous-wave (CW) mode at room temperature are displayed in Fig.~\ref{fig:LIV}. In each plot, an orange-shaded area has been added to visually show the region of operation that has been investigated in the noise measurements for each laser (e.g., the region under study for the  QCL goes from 120 to 180 mA, as visible in Fig.~\ref{fig:LIV}(a)). We see that while the slope of the QCL's LI curve is almost constant after reaching the threshold current, the corresponding value for the ICL changes at approximately 70 mA (Fig.~\ref{fig:LIV}(b)). This is in coincidence with a mode jump as described in appendix \ref{app:optical_spectra}. The range of drive current for the noise analysis is chosen such that the laser has the widest possible range of operation while maintaining a stable regime of emission during modulation.
From the LI curves, it is possible to extract the electron-to-photon conversion rate. This parameter, defined for electrically pumped lasers, is called laser quantum efficiency (QE). More precisely, the QE is the ratio between the number of generated photons and the number of pumped electrons, per stage, of the active medium. The QE at a certain operating point can be calculated as the power $P$ at a bias current $I_{\mathrm{bias}}$, multiplied by the electron charge \textit{e}, divided by the energy of the emitted light ${h \nu}$ and by the number of cascade stages $N_{\mathrm{C}}$:
\begin{equation}
    \mathrm{QE}~=~\frac{P/h \nu}{I_{\mathrm{bias}}/e} \cdot \frac{1}{N_{\mathrm{C}}}~=~\frac{P}{I_{\mathrm{bias}}} \cdot \frac{e}{h\nu} \cdot \frac{1}{N_{\mathrm{C}}}
    \label{eq:QE}
\end{equation}
A further device characterizing parameter that can be extracted from the LIV curves is the wall-plug efficiency (WPE), which is the fraction of electrical power that is converted into optical output power of the respective laser. The WPE can be calculated as the ratio between the optical power measured at the output, $P$, and the input electrical power $P_{\mathrm{el}} = V \cdot I_{\mathrm{bias}}$ where $V$ is the voltage applied to the laser chip:
\begin{equation}
    \mathrm{WPE}~=~\frac{P}{V \cdot I_{{\mathrm{bias}}}}
    \label{eq:WPE}
\end{equation}
\begin{table}[htb!]
    \centering
     \resizebox{0.35\linewidth}{!}{
    \begin{tabular}{|c|c|c|}
        \hline
        &  QCL @ 180 mA  & ICL @ 90 mA \\
        \hline
        WPE & 0.064  & 0.003  \\
        \hline
        QE & 0.095  & 0.022  \\
        \hline
        $\mathrm{N_C}$ & 25 & 6 \\
        \hline
    \end{tabular}
    }
    \caption{WPE and QE of the two lasers calculated at their maximum attainable operating points, and their number of cascades $\mathrm{N_C}$.}
    \label{tab:QE}
\end{table}
\\
Table~\ref{tab:QE} reports QE and WPE of the two lasers calculated at their maximum attainable operating points, where both the QE and the WPE are also at their maximum. From the Table, it is observed that the QCL shows better performance in terms of both WPE and QE (WPE~=~6.4~\% and QE~=~9.5~\%) than the ICL (WPE~=~0.3~\% and QE~=~2.2~\%). 

\paragraph{Detector} 
The used detector is a commercial Vigo HgCdTe detector (PVI-4TE-5-2x2) with a nominal bandwidth of \SI{175}{MHz}. The detector responsivity, i.e. the slope of the linear trend of generated photocurrent when plotted against incident power, is 1.48~A/W, leading to a quantum efficiency (ratio between number of generated electrons and number of incident photons) of 41~\%, for incident radiation at a wavelength of \SI{4.5}{\micro m}. For the same wavelength, saturation occurs when the incident power exceeds \SI{1}{mW} approximately. Finally, the detector background noise level is on the order of $\mathrm{1 \cdot 10^{-5} ~ nA^2/Hz} $. Note that the declared background noise is obtained with the detector switched on and covered. Throughout all this work, the detector was operated in its linear regime to avoid detector saturation and other disruptive effects.

\paragraph{Current driver}
The current driver used in these experiments is a custom-made low-noise current driver provided by ppqSense (QubeCL)\footnote{https://www.ppqsense.com/qube-cl/}. In particular, this device has been specifically developed to have a reduced current noise in the bias current range of the tested lasers (i.e, around 100 mA). The device capability in terms of current noise reduction has been characterized in this work as described in detail in the following section. In particular, the noise features are shown in Fig.~\ref{fig:RIN_ICL}~a and Fig.~\ref{fig:RIN_QCL}~a.

\subsection{Measurement procedure and data analysis}
\subsubsection{Setup and measurements}\label{sec:setup_meas}
\begin{figure}[htb!]
\centering\includegraphics[width=0.5\linewidth, keepaspectratio]{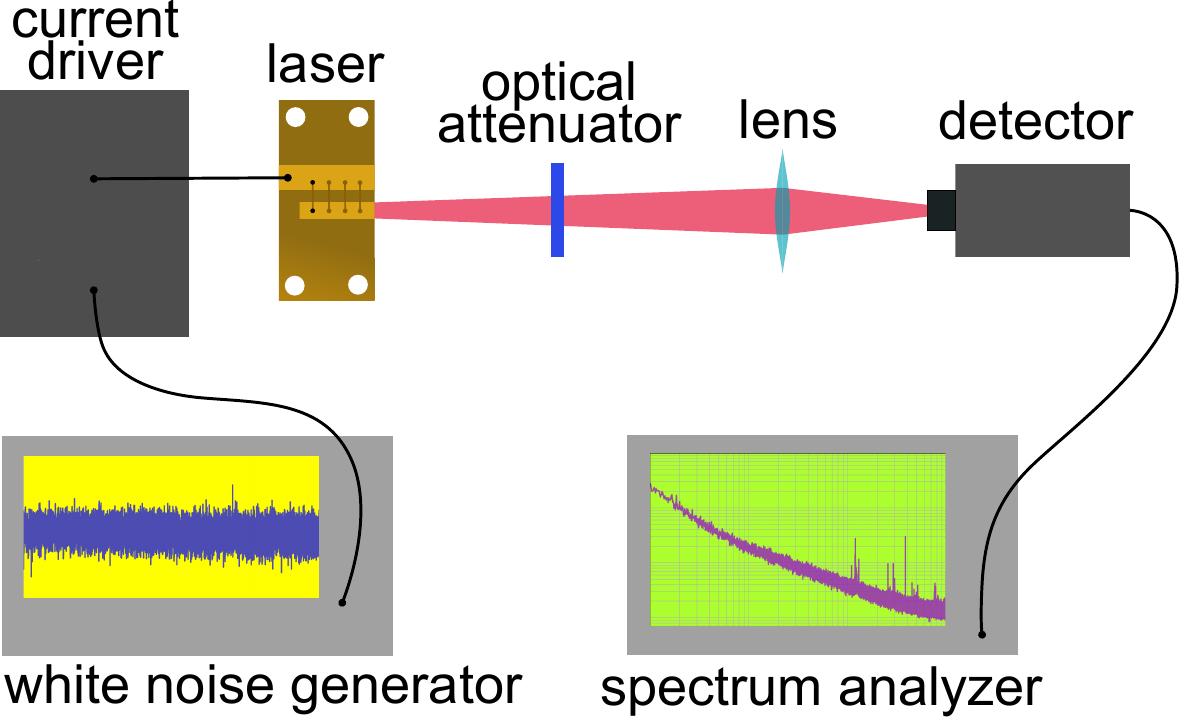}
\caption{Scheme of the setup used for the measurement of the lasers' intensity noise power spectral density (INPSD). The tested laser is supplied with the low-noise current driver. The current is modulated by means of the driver modulation unit, with white noise generated by an arbitrary function generator. While the laser's operating point is tuned across the respective orange range in the plots from Fig. \ref{fig:LIV}, for each value of bias current the peak-to-peak amplitude of the white noise is varied. The laser light is then collected onto a detector whose signal is processed by a spectrum analyzer to retrieve the wanted INPSD.}
\label{fig:setup}
\end{figure}
To study the contribution from the driver's electrical current noise to the laser's emitted intensity noise, we measure the transfer function from the bias current fluctuations $\Delta I_{\mathrm{bias}}$ to the laser optical power fluctuations $\Delta P$, using the following procedure: we drive the CW lasers with the high-performance, low-noise current driver provided by ppqSense (QubeCL). The emitted laser light is then collected onto the Vigo detector, whose photocurrent output signal is sent to a real-time spectrum analyzer (Tektronix RSA5106A). Note that, for the subsequent signal analysis, the spectra provided by the spectrum analyzer were rescaled to match a double-sided power spectral density. The driver comes with an integrated modulation unit, used to add a white noise modulation on top of the bias current. 
A scheme of the setup is shown in Fig.~\ref{fig:setup}, where a further component, namely an optical attenuator placed in front of the laser, is depicted. Attenuating the laser beam was necessary for both lasers to avoid detector saturation. \\
The laser operating point is varied by a stepwise increase of the bias current, and at each operating point, the bias current is modulated with white noise at different amplitudes. For each of these modulation amplitudes, the detector AC signal is recorded on the spectrum analyzer. The result is a set of intensity noise power spectral densities (INPSDs) for each operating point.

We want to point out that optical attenuation in front of the detector corresponds to a loss of photons from the emitted flux, which may prevent us from observing sub-Poissonian photon statistics \cite{henry_quantum_1996}.
\\
If we want to compare noise spectra from different lasers, we should take into account that each laser emits at different power. This is done by computing the relative intensity noise (RIN$_{\mathrm{laser}}$), which is a measure of the fluctuations in photocurrent $\Delta I_{\mathrm{pc}}$ per unit frequency, divided by the square of the average photocurrent $I_{\mathrm{pc}}$:
\begin{equation}
    \mathrm{RIN}_{\mathrm{laser}}=~\frac{\Delta I_{\mathrm{pc}}^2}{\mathrm{RBW} \cdot I_{\mathrm{pc}}^2}~=~\frac{\mathrm{INPSD}}{I_{\mathrm{pc}}^2}
    \label{eq:RIN_laser}
\end{equation}
where RBW is the resolution bandwidth of the instrument used for acquisition, i.e. in this case, the spectrum analyzer, and where INPSD = $\frac{\Delta I_{\mathrm{pc}}^2}{\mathrm{RBW}}$\footnote{To compute the INPSD starting from the signal in dBm acquired with the spectrum analyzer, one has to first convert from dBm to mW, then to units of $\mathrm{A/Hz^2}$ using the right conversion factors, such as the detector's transimpedance and gain. All those factors had been rigorously measured in the previous devices' characterization.}. \\
\\
\begin{figure}[htb!]
\centering\includegraphics[width=0.5\linewidth, keepaspectratio]{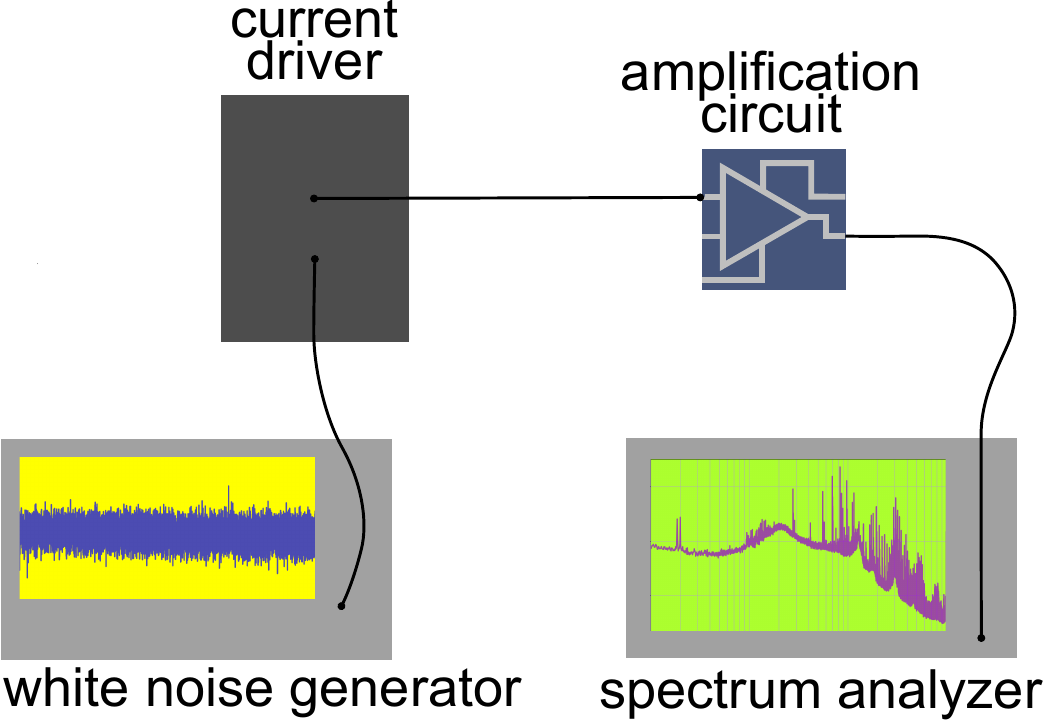}
\caption{Setup for the measurement of the driver's INPSD. The bias current is converted to a voltage signal via an home-made amplification circuit, whose core consists of two op-amps (AD797N and AD8001), and which is battery-supplied to extinguish the electric noise from the wall-plug. The amplified signal is sent to the spectrum analyzer. For each value of bias current, the latter is modulated with a white noise of varying amplitude by means of a function generator. In the same fashion, the lasers' bias current has been modulated for the measurements of the lasers' INPSD.}
\label{fig:setup_qube}
\end{figure}

To understand the impact of the photocurrent noise on the laser intensity noise in view of a possible control, we are particularly interested in comparing the RIN of the laser (RIN${_{\mathrm{laser}}}$) to the RIN of the driver (RIN${_{\mathrm{driver}}}$), in other words, the RIN of the photocurrent from the detector to the RIN of the bias current generated via the current driver. The INPSD of the bias current is directly measured by connecting the driver's output to the spectrum analyzer after proper amplification with a battery-backed amplification circuit (see Fig. \ref{fig:setup_qube}). The signal is acquired through a spectrum analyzer. This is done for all values of bias current used to supply the ICL and the QCL, i.e., from 40 to 180 mA, in 10 mA steps. At each step, the driver's bias current was modulated with a white noise of varying amplitude, the same way it was done with the lasers' bias current when measuring the lasers' INPSD. The driver's INPSD is divided by the square of the bias current $\mathrm{I_{bias}}$ to get the RIN of the driver (RIN$_{\mathrm{driver}}$), similarly to eq. \ref{eq:RIN_laser}:
\begin{equation}
    \mathrm{RIN}_{\mathrm{driver}}=~\frac{\mathrm{{INPSD}}}{ I_{\mathrm{bias}} ^2}
    \label{eq:RIN_qube}
\end{equation}

\subsubsection{Noise analysis and transfer function} \label{sec:noise_transf}

\begin{figure*}[htb!]
\centering\includegraphics[width=0.9\linewidth, keepaspectratio]{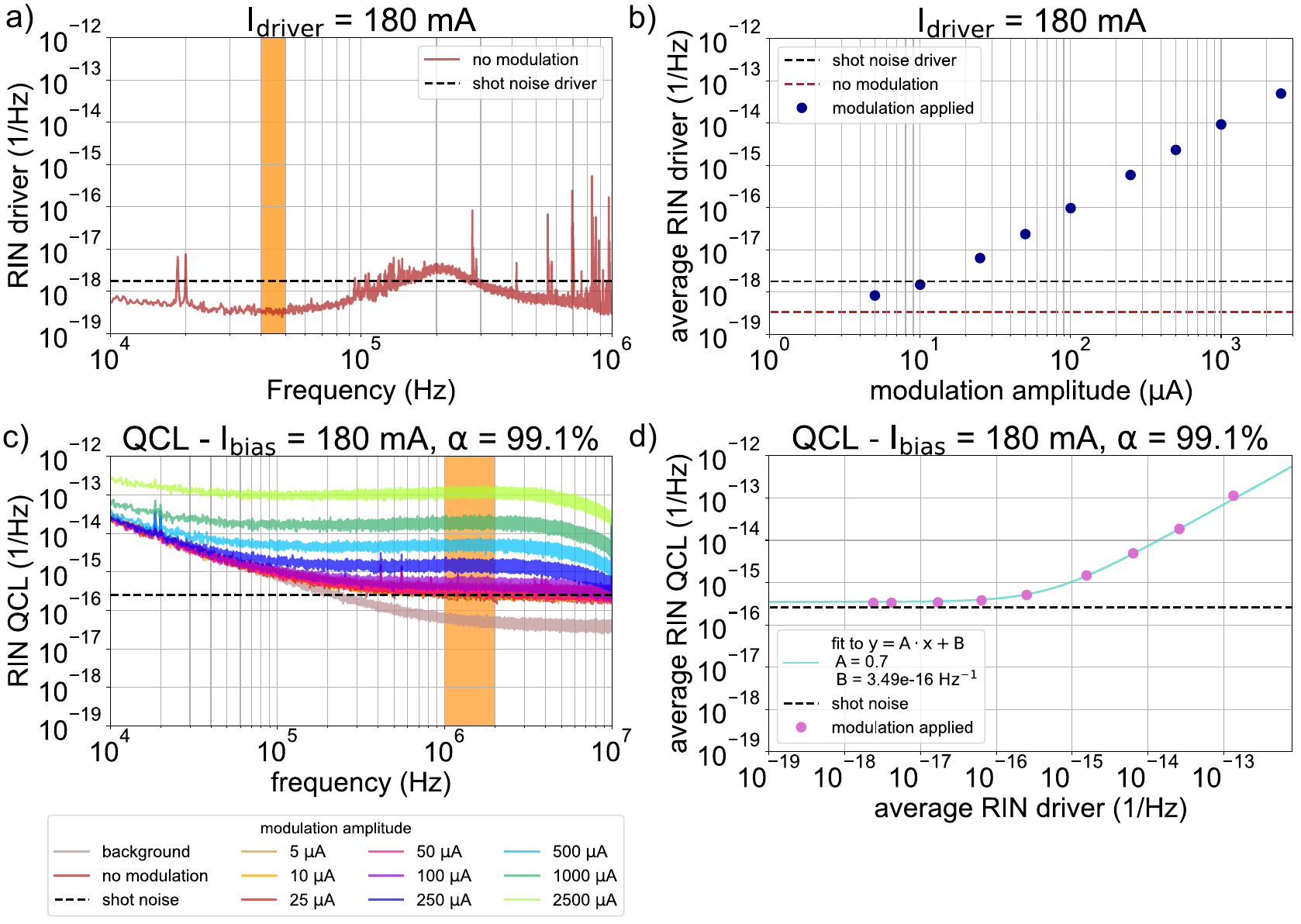}
\caption{Panel (a): RIN spectrum of the driver supplying 180 mA of bias current (RIN$_{\mathrm{driver}}$). The RIN$_{\mathrm{driver}}$ was measured for different amplitudes of applied modulation, similarly to the graphic in panel (c). But for clarity, the plot only shows the trace corresponding to the non-modulated current signal, together with the computed corresponding shot noise level symbolized by the black dashed line. The orange-shaded area indicates the frequency range where the spectra were averaged. Panel (b): Average RIN$_{\mathrm{driver}}$ plotted against the corresponding modulation amplitude. The red dashed line is the average RIN$_{\mathrm{driver}}$ of the non-modulated current signal, taken as a reference, while the average RIN of the modulated signal is represented by blue dots. The black dashed line represents the corresponding shot noise level. Panel (c): RIN spectra of the QCL pumped at 180 mA, with an attenuation of the optical power $\mathrm{\alpha}$ of 99.1\% (RIN$_{\mathrm{QCL}}$). Traces of different colors correspond to different peak-to-peak amplitudes of the white noise added to the bias current, as reported in the legend, while the term "no modulation" stands for the non-modulated signal. The old-rose line represents the detector's background, while the black dashed line is the computed laser's shot noise. The orange-shaded area shows the frequency region where the average on the RIN spectra has been performed. Panel (d): Similarly to panel (b), this plot represents the average RIN$_{\mathrm{QCL}}$ versus the average RIN$_{\mathrm{driver}}$, i.e. the transfer function from the driver's bias current to the photon flux. The transfer function is a line whose slope A and offset B are stated in the legend. }
\label{fig:RIN_QCL}
\end{figure*}

The result of the noise measurements of the lasers and current driver is a set of RIN spectra for each value of bias current. Fig.~\ref{fig:RIN_QCL}(a) reports the driver's RIN spectrum with no modulation applied, for a bias current of 180 mA, together with the driver's shot noise power spectral density ($\mathrm{PSD}_{\mathrm{SN}}=2 \cdot e \cdot I_{\mathrm{bias}}$~\cite{gabbrielli_mid-infrared_2021}). This is to highlight the fact that the bias current intensity fluctuations are below the shot noise level, i.e. they are sub-shot noise. In particular, we observe that in the frequency range highlighted by the orange-shaded area, the RIN$_{\mathrm{driver}}$ lies 7.3~dB below the shot noise (black dashed lines in Fig.~\ref{fig:RIN_QCL}(a)). The orange-shaded area indicates the frequency range where the spectra are averaged for the subsequent analysis. In fact, averaging the spectrum in the frequency range where it is most flat, and with fewer peaks or flicker-noise contributions, provides information about how the RIN evolves with increasing modulation amplitude, as visible in Fig.~\ref{fig:RIN_QCL}(b). In this graph, the average of the RIN$_{\mathrm{driver}}$ spectra at 180 mA, computed between 40 and 50 kHz, is plotted against the applied white-noise modulation amplitude in \SI{}{\micro A}. The non-modulated current signal and the driver's shot noise (red and black dashed lines, respectively) are also plotted as reference. This plot clearly shows that the non-modulated current signal is sub-shot-noise and that by adding a modulation with varying amplitude, the noise increases accordingly. Note that the increment in the RIN$_{\mathrm{driver}}$ occurs as soon as a small modulation is applied, suggesting that the driver's modulation unit responds immediately to the additional, external modulation. The same analysis is performed for the laser noise with the RIN$_{\mathrm{laser}}$ spectra. Fig.~\ref{fig:RIN_QCL}~(c) depicts the RIN spectra of the QCL (RIN$_{\mathrm{QCL}}$) driven at 180 mA, each color corresponding to a different modulation amplitude of the bias current, as reported in the legend. The plot also comprises the RIN of the non-modulated signal, as well as the detector's background noise (old-rose trace), which is the signal measured with the detector turned on and covered. The (RIN$_{\mathrm{QCL}}$) traces are well above this trace, assuring that the detector's background does not limit the measurements. Finally, the computed shot noise power spectral density (PSD) of the incident radiation, measured in terms of generated photocurrent ($I_{\mathrm{pc}}$) by the detector, is displayed as a black dashed line. This value is estimated from the detector's photocurrent as $ \mathrm{PSD_{\mathrm{SN}}}=2 \cdot e \cdot I_{\mathrm{pc}}$. The fact that the (RIN$_{\mathrm{QCL}}$) spectra at low modulation amplitudes are at the shot noise level is mainly due to the strong attenuation factor $\mathrm{\alpha}$ (99.1\%, as declared in the plot's header). For a more detailed explanation of this effect and the verification of the shot noise level, see appendix \ref{app:balanced_detector}.

For the analysis of the RIN$_{\mathrm{QCL}}$ spectra, we choose again to average them in the frequency range where they are most flat and less peaks are present, i.e. the region 1-\SI{2}{MHz}. In particular,  in Fig.~\ref{fig:RIN_QCL}(d) the pink dots show the trend of the average RIN$_{\mathrm{QCL}}$ as a function of the average RIN$_{\mathrm{driver}}$ when amplitude-modulated with the white-noise modulation. The relationship between these two RIN quantities is the transfer function from the current driver to the laser's photon flux. The trend is linear, as shown by the linear fit performed on the dataset, represented by the light-blue line. The fitting parameters A and B, representing the slope and the offset of the linear fit respectively, are reported in the legend. By comparing the transfer function with the behavior of the average RIN$_{\mathrm{driver}}$ versus modulation amplitude of Fig. \ref{fig:RIN_QCL}(b), we notice that the RIN$_{\mathrm{driver}}$ has a significant contribution to the RIN$_{\mathrm{QCL}}$ starting by a value of approximately \SI[parse-numbers=false]{10^{-16}}{Hz^{-1}}, corresponding to a modulation amplitude of \SI{100}{\micro A}. At lower values, the noise from the current driver is buried in the intrinsic noise of the QCL, which in this case is at the shot noise level because of the strong attenuation being used.

\begin{figure*}[htb!]
\centering\includegraphics[width=0.9\linewidth, keepaspectratio]{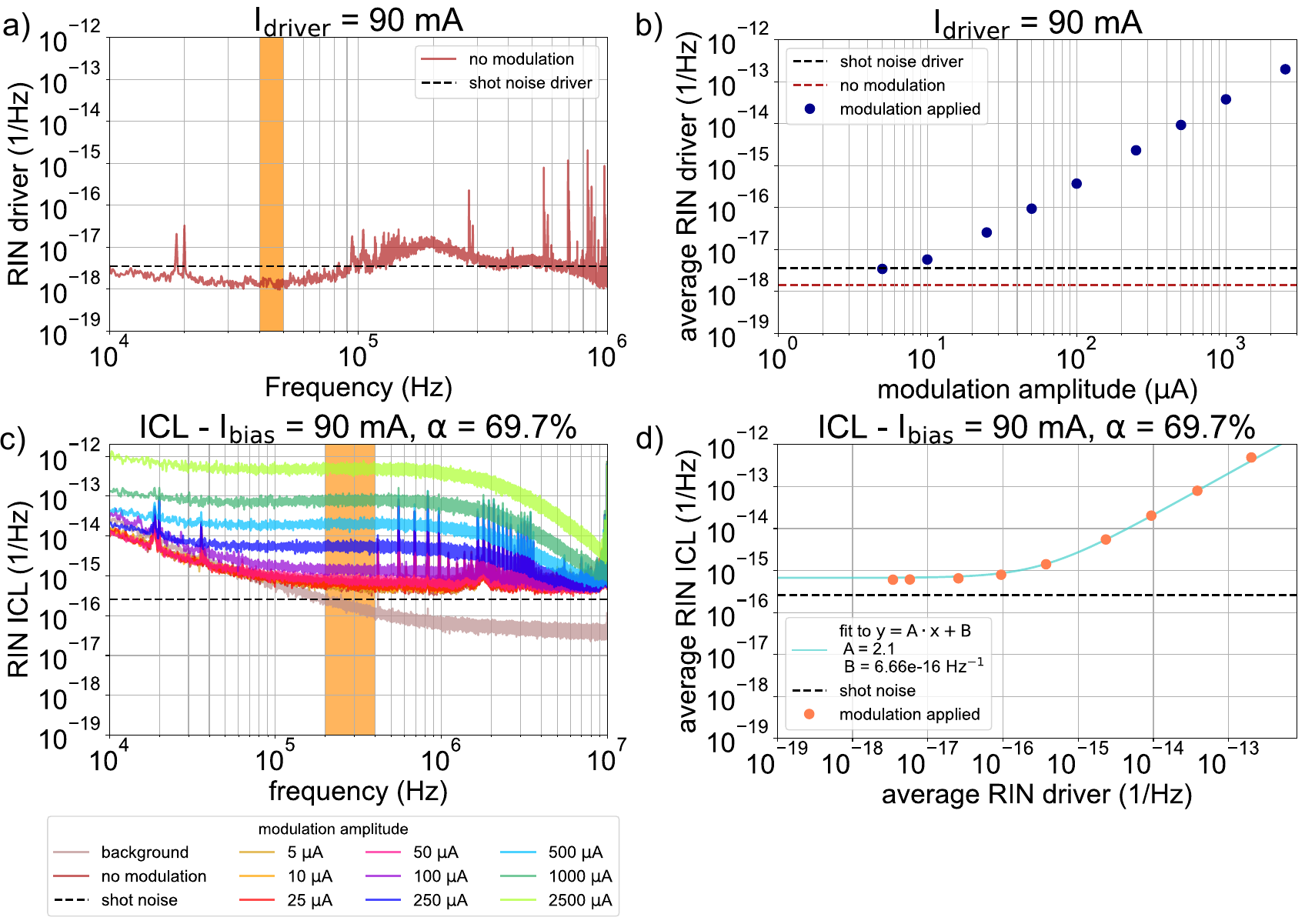}
\caption{Noise analysis for the ICL at 90 mA of bias current. The panels (a), (b), (c), and (d) are similar to those of Fig. \ref{fig:RIN_QCL}, where (a) and (b) refer to the RIN$_{\mathrm{driver}}$ at 90 mA, while (c) and (d) to the RIN$_{\mathrm{ICL}}$ when the ICL is driven at 90 mA, with an optical attenuation $\mathrm{\alpha}$ of 69.7\%. Here, the RIN$_{\mathrm{ICL}}$ has been averaged in the range between 200 and 400 kHz.}
\label{fig:RIN_ICL}
\end{figure*}

Similar results are obtained with the ICL after performing the same analysis of the RIN at each operating point (i.e. from 40 to 90 mA of bias current, in 10 mA steps). Fig. \ref{fig:RIN_ICL} displays four plots analogous to those of Fig. \ref{fig:RIN_QCL}, related to the RIN$_{\mathrm{driver}}$ at 90 mA (Fig. \ref{fig:RIN_ICL}(a) and (b)) and to the RIN of the ICL (RIN$_{\mathrm{ICL}}$) driven at 90 mA, with an optical attenuation $\mathrm{\alpha}$ of 69.7\% (Fig. \ref{fig:RIN_ICL}(c) and (d)). At this operating point, the driver is 3.9~dB below the shot noise in the range between 40 and 50 kHz, as marked by the orange-shaded area in Fig. \ref{fig:RIN_ICL}(a). The average RIN$_{\mathrm{driver}}$ reaches and overcomes the shot noise level as soon as a modulation is applied, as visible in Fig. \ref{fig:RIN_ICL}(b). The RIN spectra of the ICL shown in Fig. \ref{fig:RIN_ICL}(c) are slightly above the shot noise level. This is because the ICL emits much less power than the QCL (see Fig. \ref{fig:LIV}), therefore less attenuation was needed to stay in the detector's linear regime of responsivity. Finally, Fig. \ref{fig:RIN_ICL}(d) represents the transfer function from the driver's bias current to the ICL's emitted photon flux. Similarly to the QCL, the ICL's average RIN is significantly affected by the external modulation only starting by an amplitude of \SI{100}{\micro A}, corresponding to an average RIN$_{\mathrm{driver}}$ of $\SI[parse-numbers=false]{3.7 \cdot 10^{-16}}{Hz^{-1}}$.\\
\\
We are now interested in studying the transfer function from electric current noise to photon flux noise. The input variable is the RIN$_{\mathrm{driver}}$, and the output quantity is the measured RIN$_{\mathrm{laser}}$. This means the transfer function $T$ relates the normalized bias current fluctuations ($\Delta I_{\mathrm{bias}}^2 / I_{\mathrm{bias}}^2$) to the normalized output power fluctuations ($\Delta P^2 /P^2$), which in turn are directly proportional to the photocurrent fluctuations ($\Delta I_{\mathrm{pc}}^2 / I_{\mathrm{pc}}^2 $) of the detector operated in the linear responsivity regime:
\begin{equation}
    T: ~~ \frac{\Delta I_{\mathrm{bias}}^2}{\bigl (I_{\mathrm{bias}}\bigr )^2} \longrightarrow \frac{\Delta I_{\mathrm{pc}}^2}{I_{\mathrm{pc}}^2} 
    \label{eq:transfer_fct}
\end{equation}
In practice, we want to study the trend of the parameter $A$, i.e. the slope obtained from the fit procedure shown in Figs.~\ref{fig:RIN_QCL}(d) and \ref{fig:RIN_ICL}(d), by varying the bias current. Both for the QCL and the ICL, the transfer function $\mathrm{T:~ x \longrightarrow y}$ (Eq. \eqref{eq:transfer_fct}) was observed to follow a linear trend, and was thus fitted to a line: $y = A \cdot x + B$. We define parameter $A$ as the \textit{transfer coefficient}, which is a measure of the laser's sensitivity to the current fluctuations. On the other hand, $B$ represents the initial offset, which is the lasers' average RIN when no modulation is applied. 
Regarding the transfer coefficient $A$, it is interesting to see its evolution in relation to the bias current. The transfer coefficient at each operating point is extrapolated as a fit parameter from the linear regression on the average RIN$_{\mathrm{laser}}$ (as reported in the legends of Figs. \ref{fig:RIN_QCL}(d) and \ref{fig:RIN_ICL}(d)). Specifically, $A$ is the slope of the transfer function, defined as the ratio between the increment in $y$ and the increment in $x$. The variables y and x are the RIN$_{\mathrm{laser}}$ and the RIN$_{\mathrm{driver}}$, respectively, as stated in Eq. \eqref{eq:transfer_coeff_I_bias}. The detector's photocurrent is proportional to the laser power, therefore, it can be regarded as being proportional to the bias current $\mathrm{I_{bias}}$ minus the laser's threshold current $\mathrm{I_{th}}$. As a result, we have the following dependence of the transfer coefficient $A$ on the bias current: 
\begin{equation}
    A = \frac{y-B}{x} = \frac{\Delta I_\mathrm{pc}^2 / \bigl (I_\mathrm{pc}\bigr )^2}{\Delta I_\mathrm{bias}^2 / \bigl (I_\mathrm{bias}\bigr) ^2} + b \approx  a \cdot \frac{(I_\mathrm{bias}\bigr )^2}{\bigl (I_\mathrm{bias}-I_\mathrm{th}\bigr )^2} + b 
    \label{eq:transfer_coeff_I_bias}
\end{equation}

\begin{figure}[htb!]
\centering\includegraphics[width=0.6\linewidth, keepaspectratio]{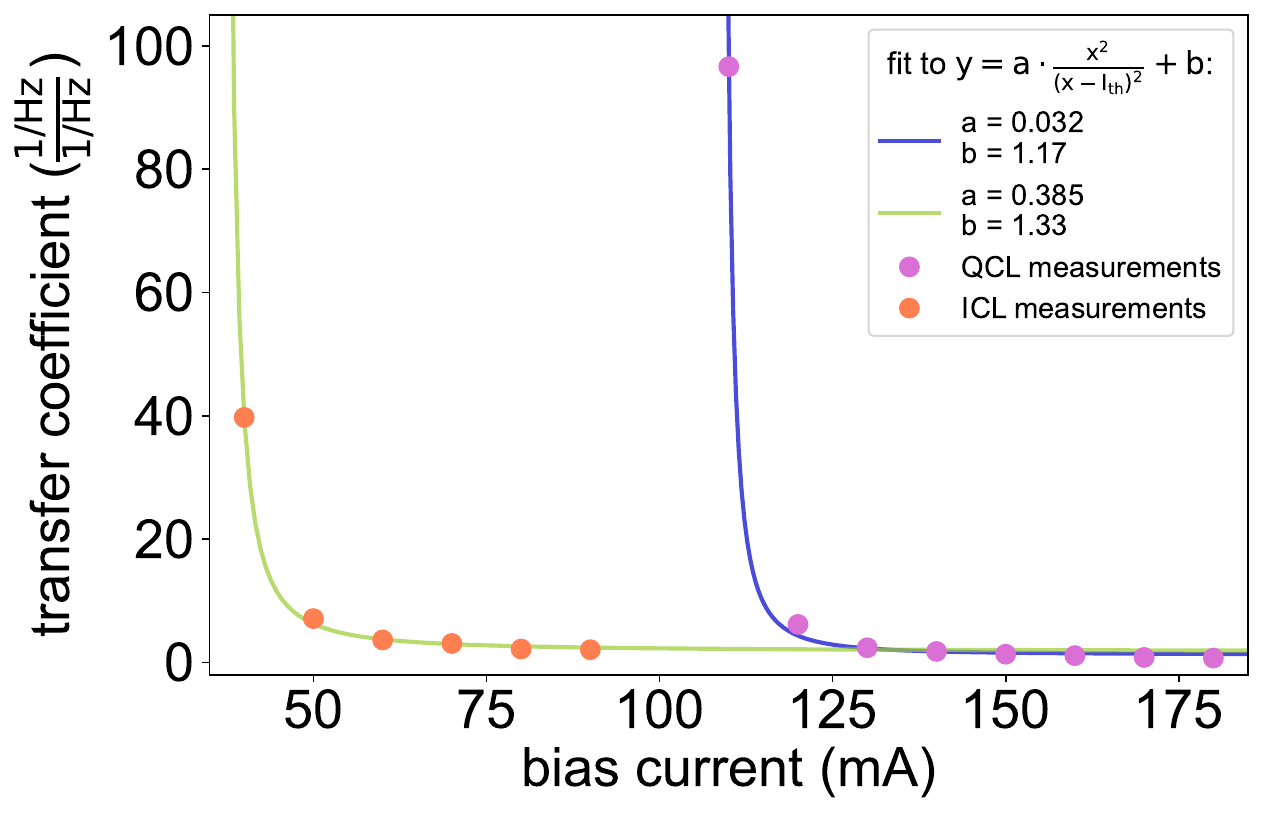}
\caption{Plot of the transfer coefficient $A$ as a function of the bias current for the ICL (orange dots) and the QCL (pink dots). The green and blue lines represent the fits performed according to the equation given in the legend, whose fitting parameters $a$ and $b$ are also given in the legend for each fit.}
\label{fig:transf_coeff} 
\end{figure}

The relationship between the transfer coefficient $A$ and the bias current as given by Eq. \eqref{eq:transfer_coeff_I_bias} is highlighted in Fig. \ref{fig:transf_coeff}, where the transfer coefficient of the two lasers is displayed as a function of $\mathrm{I_{bias}}$, and fitted with the model $y = a \cdot x^2/ \bigl (x-I_{\mathrm{th}} \bigr) ^2 + b$. We point out that this relation holds as far as $a$ and $b$ can be considered as constant parameters and the curve does not diverge. The first condition happens when the ratio between the absolute fluctuations $\Delta I_\mathrm{pc}^2 / \Delta I_\mathrm{bias}^2 $ is constant and when the variation of $B/x$ is negligible with respect to that of the quantity $y/x$. The second condition breaks when $I_\mathrm{bias}$ approaches $I_\mathrm{th}$ and the function diverges. 
As a consequence, the transfer coefficient has its peak value near to laser threshold, and rapidly decays towards $a+b$ with increasing bias current, as shown by the values of parameter b in the legend of Fig. \ref{fig:transf_coeff}, which is approximately 1.2 for the QCL and 1.7 for the ICL. This fact demonstrates that the laser above threshold reaches a stable regime of operation not only in terms of emitted power but also in terms of power fluctuations. 

\subsection{Discussion}

For this specific application, we would like the transfer coefficient to be as high as possible, meaning that the laser has a high sensitivity to the pump current fluctuations. This will in turn result in an immediate reduction of noise when the pump current fluctuations are further suppressed. Taking a glance at Fig.~\ref{fig:transf_coeff}, where the transfer coefficient of the two lasers is plotted against the respective values of drive current, it is evident that the transfer coefficient is highest near threshold. However, this is mainly related to the fact that the lasers' RIN is also at its highest near threshold (ideally when the current $\rightarrow 0$, the RIN$\rightarrow \infty$). As already pointed out in the discussion of eq. \ref{eq:transfer_coeff_I_bias}, the data analysis holds as long as we work relatively far from threshold.  The transfer coefficient's trend tends towards an asymptotic stabilization when we move away from threshold. In detail, comparing the two setpoints reached, the ICL has a higher value of b (i.e. 1.33 for ICL vs. 1.17 for the QCL), indicating a higher sensitivity in terms of noise transfer. More generally, this analysis offers a comprehensive description of classical noise transfer from laser bias current to the emitted output power. In addition, this work demonstrates that the $\mathrm{RIN}_\mathrm{laser}$ is insensitive to small bias current fluctuations, as can be deduced from Figs. \ref{fig:RIN_ICL}(d) and \ref{fig:RIN_QCL}(d).
This fact suggests that by driving state-of-the-art MIR lasers with a sub-shot-noise bias current, the laser's intensity noise cannot be significantly reduced. 
\begin{table}[htb!]
    \centering
     \resizebox{0.6\linewidth}{!}{
    \begin{tabular}{|c|c|c|c|}
        \hline
        device & reference & QE (laser + detector) & amount of squeezing \\
        \hline
        LED & \cite{yamamoto_photon_1992} & 48\% & 14 dB \\
        \hline
        LED & \cite{giacobino_quantum_1996,marin_squeezing_1995} & 66\% (laser only) & 1.4 dB \\
        \hline
        VCSEL & \cite{kaiser_amplitude-squeezed_2001} & 20\% & 0.4 dB \\
        \hline
        quantum dot laser & \cite{ding_observation_2024} & 86\% & 1.7 dB \\
        \hline
        ICL @ 90 mA & this work & 0.9\% & \\
        \hline
        QCL @ 180 mA & this work & 3.9\% & \\
        \hline
    \end{tabular}
    }
    \caption{Overview of the performances from other works regarding the generation of sub-shot noise light by lasers backed by a quiet pump. The overall QE comprises both the laser and the detector's QE. The amount of squeezing reached by the reference works relates to how far below shot noise the measured intensity noise of the laser is. }
    \label{tab:overview}
\end{table}
From a qualitative point of view, we expected the transfer coefficient to be directly proportional to the QE of the laser. As a matter of fact, a higher rate of electron-to-photon conversion is directly related to a higher probability of transfer from electron-to-photon statistics. This trend would be visible if the QE of the tested lasers was higher, as reported in the mentioned works \cite{yamamoto_photon_1992,giacobino_quantum_1996,marin_squeezing_1995,kaiser_amplitude-squeezed_2001,ding_observation_2024}. In the present work, however, the QE of the lasers is limited to about 1\% or below for all of the tested devices. An overview of our performance compared to that of other representative works is shown in Table \ref{tab:overview}.
Therefore, on the pathway towards controlling the intensity noise of mid-infrared cascade lasers (QCLs and ICLs) at the quantum level by acting on the bias current electron statistics, the QE of the lasers should be improved, which involves an optimization of the design and fabrication process of such devices.
 
\section{Conclusions}
In this work, we investigated the electron-to-photon noise transfer in mid-infrared semiconductor lasers when driven via a custom-made, sub-shot-noise current source. In particular, we compared how two different types of single-mode MIR lasers, namely a QCL and an ICL, convert the injected electric bias current noise into intensity noise of the emitted light. The analysis was based on the characterization of key parameters such as quantum efficiencies, wall-plug efficiencies, and noise transfer coefficients, which play a crucial role in controlling the intensity noise of the laser when acting on the driving current.   \\
The results show in both cases that the transfer of noise, evaluated in terms of the transfer coefficient, rapidly reaches a setpoint as far as the bias current deviates from the threshold value. The measured trends suggest that the tested ICL is more responsive to bias fluctuations than the QCL. We point out that the two selected lasers are state-of-the-art, commercially available devices. Therefore, this work offers, on the one hand, a tested methodology of characterization applicable to a wider set of MIR laser sources. On the other hand, it points out the limits of these devices when noise is reduced via an active control of the injected bias current. 

In general, an efficient noise transfer from an electric current to optical power fluctuations is desired, as this means that by further decreasing the current noise, we could, in principle, generate sub-shot noise light. Comparing the two devices under test with lasers from other works, emitting in the near-infrared, we clearly see a significant gap in terms of quantum efficiency, which needs to be filled to proceed in the same direction with MIR lasers. This work succeeds, however, in assessing the level of intrinsic laser noise, i.e. noise that is independent from the current driver's contribution, as far as the bias current noise gives a negligible contribution to the intrinsic laser noise. The path towards control at the quantum level of the intensity noise of MIR cascade lasers by acting on the bias current electron statistics should include optimization of laser design and fabrication strategies. In this regard, a dedicated theoretical model, describing the connection between carrier transport and photon emission at the quantum level~\cite{trombettoni_quantum_2021} and highlighting the key parameters to be further optimized for increasing the laser quantum efficiency, will be highly beneficial. This optimization should take into account a good match between the laser and the detection system in terms of power and bandwidth, meaning that the emitted laser power should be consistent with the detector saturation level, and the detector bandwidth should be adjustable according to the desired range of analysis.

\comment{
\section*{Supplementary Material}
See the supplementary material for a detailed description of the ICL and the QCL, plots and description of their optical spectra, and for a full description of the technique of balanced detection used to calibrate the RIN measurements made with the spectrum analyzer.
}

\begin{acknowledgments}
The authors acknowledge financial support by the European Union’s NextGenerationEU Programme with the I-PHOQS Infrastructure [IR0000016, ID D2B8D520, CUP B53C22001750006] "Integrated infrastructure initiative in Photonic and Quantum Sciences'', by the European Union’s Research and Innovation Programme Horizon Europe with the Laserlab-Europe Project [G.A.~n.~871124] and the MUQUABIS Project [G.A.~n.~101070546] "Multiscale quantum bio-imaging and spectroscopy", by the European Union’s QuantERA II [G.A.~n.~101017733] -- QATACOMB Project "Quantum correlations in terahertz QCL combs'', and by the Italian ESFRI Roadmap (Extreme Light Infrastructure -- ELI Project), and by the Italian Ministero dell'Università e della Ricerca (project PRIN-2022KH2KMT QUAQK), and by ASI and CNR under the Joint Project “Laboratori congiunti ASI-CNR nel settore delle Quantum Technologies (QASINO)” (Accordo Attuativo n. 2023-47-HH.0). The authors are grateful for financial support from the Austrian Research Promotion Agency (FFG) through the project ATMO-SENSE [G.A. n. 1516332].
\end{acknowledgments}

\section*{Disclosures}
The authors declare no conflicts of interest.

\section*{Data Availability}
Data underlying the results presented in this paper are available from the corresponding author upon reasonable request.

\appendix
\section{\label{app:lasers}Laser characteristics}
\subsection{QCL:}
\begin{description}
    \item[production:] Hamamatsu
    \item[emission wavelength:] around \SI{4.57}{\micro m}
    \item[type of waveguide:] ridge waveguide with top DFB grating, in buried heterostructure geometry.
    \item[waveguide dimensions:] cavity length = \SI{2}{mm}
    \item[design, materials:] active region consisting of InGaAs/InAlAs quantum wells for a total of 25 stages and with a thickness of \SI{1.325}{\micro m}. The core is sandwiched between two InP-based cladding layers grown on an InP substrate. The outcoupling facet is uncoated and has a reflectivity of 30~\%, back facet is HR coated to obtain a reflectivity of above 98~\%. 
\end{description}
\subsection{ICL:}
\begin{description}
    \item[production:] Nanoplus
    \item[emission wavelength:] around \SI{4.5}{\micro m}
    \item[type of waveguide:] ridge with top DFB grating
    \item[waveguide dimensions:] cavity length = \SI{900}{\micro m}
    \item[design, materials:] active region consisting of W-quantum well sequence: \\
    \SI{2.50}{nm} AlSb/\SI{2.22}{nm} InAs/\\\SI{2.50}{nm} \ce{In_{0.35}Ga_{0.65}Sb}/\SI{1.83}{nm} InAs/\SI{1.0}{nm} AlSb\\
    for a total of 6 stages. The active region is enclosed by  two \SI{400}{nm}-thick \ce{GaSb} separate confinement layers. Lower and upper \ce{InAs}/{AlSb} superlattice cladding layers have a thickness of 3.5 and \SI{2.0}{\micro m}, respectively.
\end{description}

\section{\label{app:optical_spectra}Optical spectra}
\begin{figure*}[htb!]
\centering\includegraphics[width=0.8\linewidth,keepaspectratio]{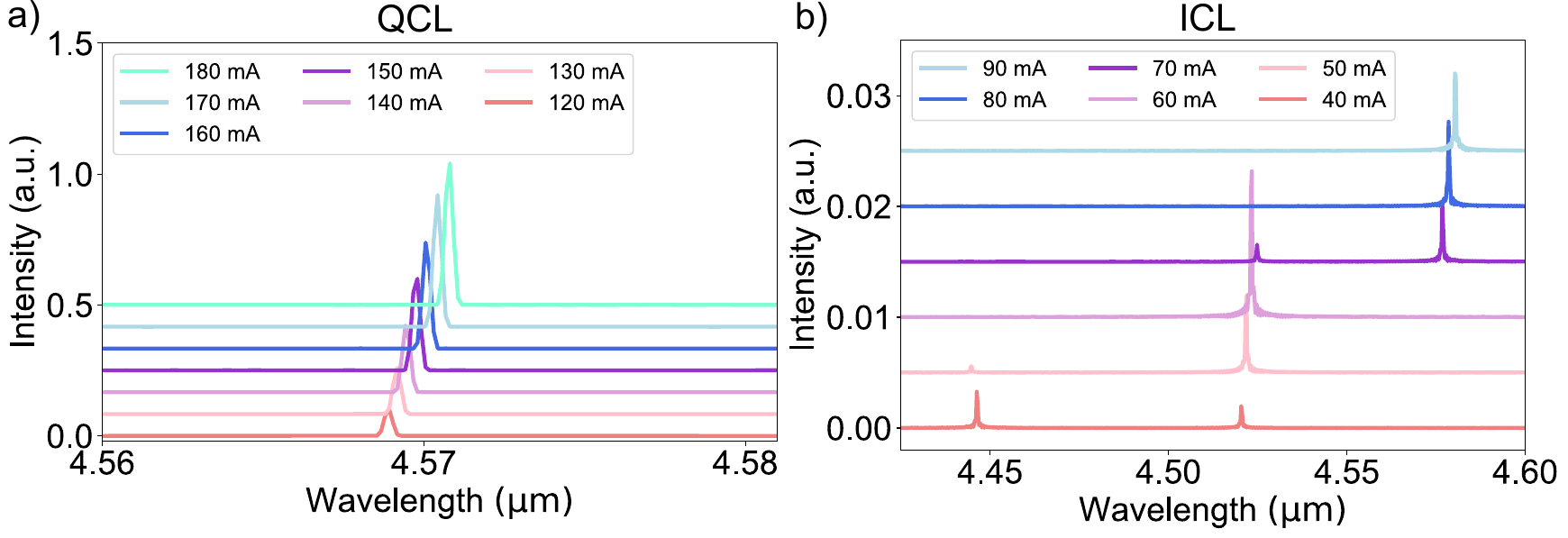}
\caption{Optical spectra of the two lasers at the values of drive current and temperature used in the RIN analysis. (a) refers to the QCL at temperature T = $\mathrm{20~\degree C}$, (b) to the DFB ICL at T = $\SI{20}{\celsius}$.}
\label{fig:OS}
\end{figure*}
The optical spectra of the two lasers are measured via an optical spectrum analyzer (Bristol, 721A and 721B), for different values of laser drive current and temperature. Fig. \ref{fig:OS} reports the optical spectra of the lasers at the values of temperature and current that have been eventually used for the noise measurements. 
Both the QCL and ICL were operated at a temperature of 20°C. 
In both plots, we observe the typical frequency tuning with drive current occurring in most semiconductor lasers. This is mostly due to variations in carrier density inside of the layers, modifying the materials' properties such as the refractive index and thus the path length. To a minor extent, increasing current also affects temperature and therefore the cavity length. The spectra clearly show single-mode emission for the QCL (Fig. \ref{fig:OS}(a)) at the peak wavelength of around \SI{4.75}{\micro m}. The spectra of the DFB ICL in Fig. \ref{fig:OS}(b) present two mode jumps, characterized by bicolour emission at two operating points, namely 40 mA and 70 mA, as mentioned in \cite{gabbrielli_shot-noise-limited_2025}. In both cases, however, there is a dominant peak that can be regarded as the laser's emission frequency. This is demonstrated by the common mode rejection ratio (CMRR), i.e. the ratio between the major peak's intensity and the secondary peak's intensity, this being 1.6 at 40 mA and 3.3 at 70 mA. As a result, the DFB ICL can be considered eligible for the second part of the work.

\section{\label{app:balanced_detector}Comparison with balanced detection measurements}
\begin{figure}[htb!]
\centering\includegraphics[width=0.5\linewidth, keepaspectratio]{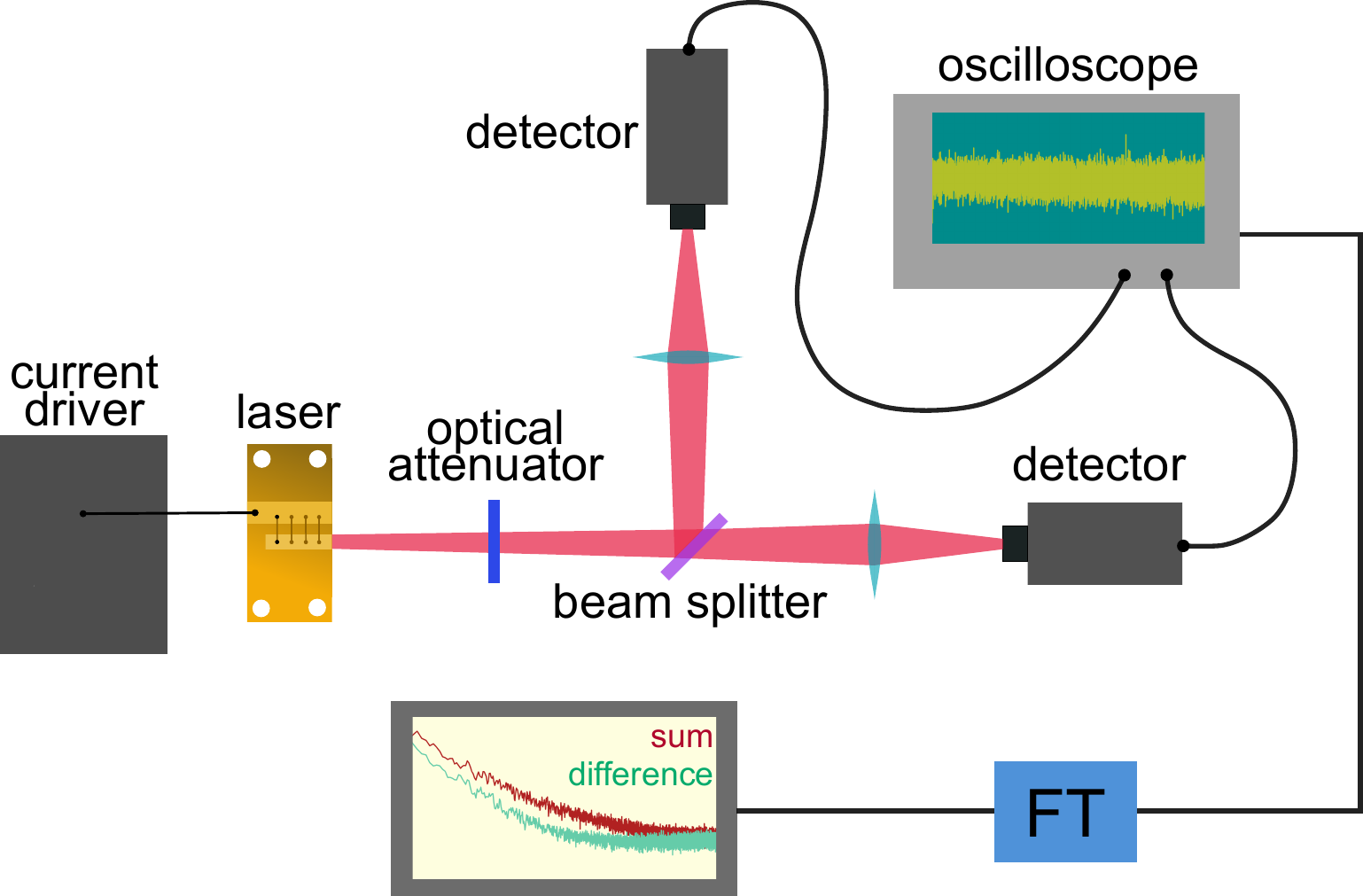}
\caption{Balanced detector setup. The attenuated MIR radiation is equally split onto two equal photovoltaic detectors by means of a 50/50 beam splitter. The resulting photocurrent signals are acquired simultaneously in the time domain via a fast oscilloscope. The Fourier transform (FT) of the sum and the difference of these signals are then computed to obtain the corresponding INPSD.}
\label{fig:setup_BD}
\end{figure}
\begin{figure*}[htb!]
\centering\includegraphics[width=0.9\linewidth, keepaspectratio]{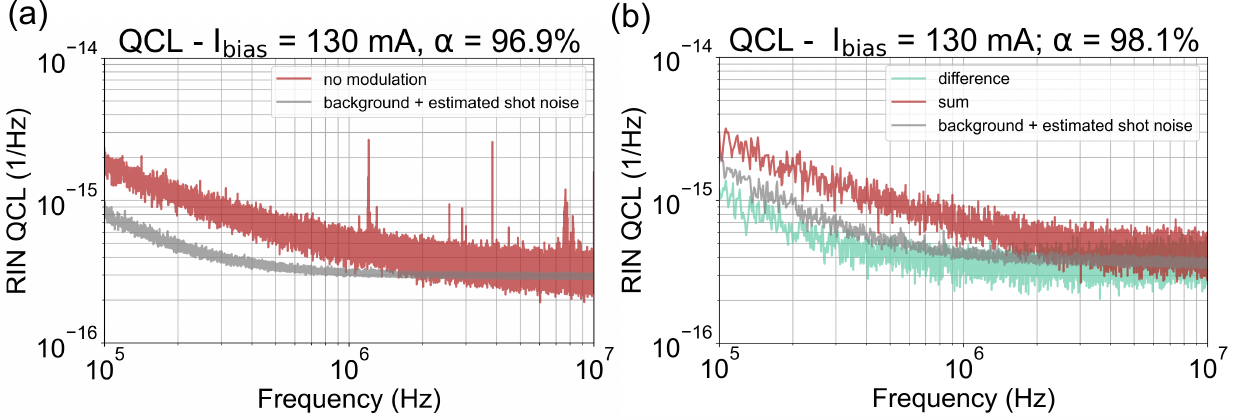}
\caption{Panel (a): RIN spectrum of the non-modulated QCL pumped at a bias current of 130 mA (red trace) and sum of corresponding computed shot noise and the detectors' background (grey trace). Panel (b): RIN spectra of the QCL at the same value of bias current, measured with the balance detector depicted in Fig.~\ref{fig:setup_BD}. In particular, the red trace is the RIN of the sum of the two signals from the two arms of the balanced detector, representing the total laser RIN, while the green trace is the difference between the two signals from the two arms, representing the shot noise. We see that the difference (green trace) approximates the grey trace, which represents the sum of shot noise (computed from the sum of the two detectors' photocurrents) and detectors' background , meaning that the system is properly calibrated. 
}
\label{fig:RIN_TF_BD_2plots}
\end{figure*}

Attenuation alters the light photon-number statistics towards a Poissonian distribution on the detector, in which case the INPSD coincides with the shot noise~\cite{fox_photon_2006}. To verify that the computed level of shot noise is correct, we performed additional measurements with a shot-noise-limited balanced detector, which had previously been calibrated~\cite{gabbrielli_mid-infrared_2021}. Given a certain radiation power, balanced detection allows for absolute measurement of the corresponding shot noise level through the computation of the difference between the two signals from the two balanced detector's arms. In our case, the MIR radiation, obtained by attenuating the laser source i.e. the QCL,  is split via a 50/50 beamsplitter, and the two parts are sent to two identical detectors (see Fig. \ref{fig:setup_BD}). The detectors' signals are acquired simultaneously with an oscilloscope and processed to find the Fourier transform (FT) of the sum and the difference between them. The sum represents the total INPSD of the incident light, while by computing the difference, the excess of common noise is canceled out and one is left with the shot noise power spectral density (this is possible if the CMRR between the excess of noise and the expected shot-noise level does not exceed the limit tolerated by the assembled setup which is in our case Fourier-frequency dependent and up to a maximum of \SI{40}{dB}~\cite{gabbrielli_mid-infrared_2021}). Following the procedure described in the main text, the INPSD is then converted into RIN, by renormalizing it to the square of the sum of the average photocurrents generated via the two detectors. 
The described technique was used to acquire the data displayed in Fig.~\ref{fig:RIN_TF_BD_2plots}(b).
This plot depicts the QCL's RIN measured at $I_\mathrm{bias}= \SI{130}{mA}$ as reported in the header, with an optical attenuation $\mathrm{\alpha}$ of 98.1\%. The red trace represents the RIN of the sum of the two signals, the green trace is the FT of the difference between the two signals, and the grey trace is the FT of the sum of the two detectors' backgrounds, plus the shot noise as evaluated from the detectors' photocurrent. By looking at Fig. \ref{fig:RIN_TF_BD_2plots}(b) we can state that the measured shot noise (i.e. the green trace) is fairly approximated by the computed one (grey trace), as the green trace approaches the estimated level of shot noise. Again, the red traces representing the total QCL's RIN (RIN$_\mathrm{laser}$) approach the shot noise because of the strong attenuation. What we also deduce from the balance detection measurements is the accuracy of our calculation of the lasers' RIN, more precisely, of the constants and conversion parameters used to calculate the RIN from the signal acquired via the spectrum analyzer. In fact, the balanced detection measurements are independently acquired in time domain with the oscilloscope, and the FT is performed downstream of the signal conversion. In other words, these are absolute measurements and can be used to calibrate the measurements performed with the spectrum analyzer. By comparing the red traces of Fig. \ref{fig:RIN_TF_BD_2plots}(a) and (b), taken at equal bias current, we can conclude that they are in good agreement with each other.

\bibliography{library}

\end{document}